# Thepost-HERAera:briefreviewoffuturelepton-ha     dronand photon-hadroncolliders

S.Sultansoy

DeutschesElektronen-SynchrotronDESY,Hamburg,Ger   many
DepartmentofPhysics,FacultyofSciences,Ankara     University,Turkey
InstituteofPhysics,AcademyofSciences,Baku,Az   erbaijan

**Abstract**

Optionsforfuture $lp,\ lA,\ \gamma p,\ \gamma A$ andFEL $\gamma A$ collidersarediscussed.

**CONTENTS**

## 1. Introduction

It is known that lepton-hadron collisions have been playing a crucial role in exploration of deep inside of Matter. For example, the quark-parton model was originated from investigation of electron-nucleon scattering. HERA has opened a new era in this field extending the kinematics region by two orders both in high $Q^2$ and small $x$ compared to fixed target experiments. However, the region of sufficiently small $x$ and simultaneously high $Q^2 (\geq 10 \text{GeV}^2)$, where saturation of parton densities should mani fest itself, is currently not achievable. It seems possi ble that $eA$ option of HERA will give opportunity to observe such phenomena. Then, the ac celeration of polarized protons in HERA could provide clear information on nucleon spi n origin.

The investigation of physics phenomena at ext remes mall $x$ but sufficiently high $Q^2$ is very important for understanding the nature of stro ng interactions at all levels from nucleus to partons. At the same time, the results f rom lepton-hadron colliders are necessary for adequate interpretation of physics at future hadron colliders. Today, linac-ring type $ep$ machines seem to be the main way to TeV scale in l epton-hadron collisions; however, it is possible that in future $\mu p$ machines can be added depending on solutions of principal issues of basic $\mu^+\mu^-$ colliders.

The aim of this brief review is to draw the a ttention of the HEP community to these facilities.

## 2. First stage: TESLA $\otimes$ HERA, LEP $\otimes$ LHC and $\mu-ring \otimes$ TEVATRON

### 2.1. TESLA $\otimes$ HERA complex

Construction of future lepton linacs tangentially t o hadron rings (HERA, Tevatron or LHC) will provide a number of additional opportunit ies to investigate lepton-hadron and photon-hadron interactions at TeV scale (see [1-3] and references therein). For example:

$$TESLA \otimes HERA = TESLA \oplus HERA$$
$$\oplus \text{TeV scale } ep \text{ collider}$$
$$\oplus \text{TeV scale } \gamma p \text{ collider}$$
$$\oplus \ eA \text{ collider}$$
$$\oplus \ \gamma A \text{ collider}$$
$$\oplus \text{FEL } \gamma A \text{ collider.}$$

The future options to collide electron and photon b eams from TESLA with proton and nucleus beams in HERA were taken into account b y choosing the direction of TESLA tangential to HERA [4-6]. It should be noted that $\gamma p$ and $\gamma A$ options are unique features of linac-ring type machines and can not be realized at LEP $\otimes$ LHC, which is comparable with TESLA $\otimes$ HERA $ep$ and $eA$ options.

### i) $ep$ option [7-10]

There are a number of reasons favoring a supercondu cting linear collider (TESLA) as a source of $e$-beam for linac-ring colliders. First of all spacin g between bunches in warm linacs, which is of the order of $ns$ (see Table I) [11], doesn't match with the bunch spacing in the HERA, TEVATRON and LHC (see Table II ) [12]. Also the pulse length this much shorter than the ring circumference. In the case of TESLA, which use standing wave cavities, one can use both shoulders in order to double electron beam energy, whereas in the case of conventional linear collider sonecanuseonlyhalfofthemachine, because the travelling wave structures can accelera te only in one direction.

The most transparent expression for the luminosity of this collider is[8]:

$$L_{ep} = \frac{1}{4\pi} \cdot \frac{P_e}{E_e} \cdot \frac{n_p}{\varepsilon_p^{\ N}} \cdot \frac{\gamma_p}{\beta_p^{\ *}}$$

for round, transversely matched beams. Using the va lues of upgraded parameters of TESLA electron beam from Table III and HERA proton beam from Table IV, we obtain $L_{ep}$=1.3·10$^{31}$cm$^{-2}$s$^{-1}$forE $_e$=300GeVoption.

The lower limit on $\beta_p^{\ *}$, which is given by proton bunch length, can be ove rcome by applying a "dynamic" focusing scheme[9], where the proton bunch waist travels with electron bunch during collision. In this scheme $\beta_p^{\ *}$islimited,inprinciple,bytheelectron bunchlength,whichistwoordersmagnitudesmaller .Moreconservatively,anupgradeof theluminositybyafactor3-4maybepossible.The refore,luminosityvaluesexceeding 10$^{31}$cm$^{-2}$s$^{-1}$seemstobeachievableforallthreeoptionsgiven inTableIII.

Further increasing of luminosity can be achie ved by increasing the number of protons in bunch and/or decreasing of normalized emittance. This requires the application of effectivecoolingmethodsatinjectorstages.Moreo ver,coolingintheHERAringmaybe necessary in order to compensate emittance growth d ue to intra-beam scattering[10]. First studies of the beam optics in the inter action region are presented in[7]and[9], where head-on collisions are assumed. The basic con cept consists of common focusing elements for both the electron and the proton beams and separating the beams outside of the low-beta insertion. However, collisions with sm all crossing angle ( ≈100 $\mu rad$) are also a matter of interest, especially for $\gamma p$and $\gamma A$options(seebelow).Furtherworkonthe subject,includingthedetectoraspects,isveryim portant.

In principle, TESLA $\otimes$HERA based $ep$ collider will extend the HERA kinematics region by an order in both $Q^2$ and $x$ and, therefore, the parton saturation regime can b e achieved.AbriefaccountofsomeSMphysicstopics (structurefunctions,hadronicfinal states, high $Q^2$ region etc.) is presented in[13], where a possibl e upgrade of the H1 detectorisconsidered.TheBSMsearchcapacityof themachinewillbedefinedbyfuture resultsfromLHC.Ifthefirstfamilyleptoquarksa nd/orleptogluonshavemasseslessthan 1 TeV they will be produced copiously (for coupling s of order of $\alpha_{em}$). The indirect manifestation of new gauge bosons may also be a mat ter of interest. In general, the physics search program of the machine is a direct e xtension of the HERA search program.

### ii) $\gamma p$option [14-16]

Earlier, the idea of using high energy photon beams obtained by Compton backscattering of laser light off a beam of high en ergy electrons was considered for $\gamma e$ and $\gamma\gamma$colliders(see[17]andreferencestherein).Then thesamemethodwasproposedin [14] for constructing $\gamma p$ colliders on the base of linac-ring type $ep$ machines. Rough estimationsofthemainparametersof $\gamma p$collisionsaregivenin[15].Thedependenceof these parameters on the distance $z$ between conversion region and collision point was analyzed in[16], where some design problems were c onsidered.

Referring for details to [16] let me note that $L_{\gamma p}$=2.6·10$^{31}$cm$^{-2}$s$^{-1}$ at $z$=10 $cm$ for TESLA⊗HERA based $\gamma p$ collider with 300 GeV energy electrons beam. Then, the luminosity slowly decreases with the increasing $z$ (factor~1/2 at $z$=10 $m$) and opposite helicity values for laser and electron beams are advantageous. Additionally, a better monochromatization of high-energy photons seen by proton bunch can be achieved by increasing the distance $z$. Finally, let me remind you that an upgrade of the luminosity by a factor 3-4 may be possible by applying a "dynamic" focusing scheme.

The scheme with non-zero crossing angle and electron beam deflection considered in [16] for $\gamma p$ option lead to problems due to intensive synchrotron radiation of bending electrons and necessity to avoid the passing of electron beam from the proton beam focusing quadrupoles. Alternatively, one can assume head-on-collisions (see above) and exclude deflection of electrons after conversion. In this case residual electron beam will collide with proton beam together with high-energy $\gamma$ beam, but because of larger cross-section of $\gamma p$ interaction the background resulting from $ep$ collisions may be neglected.

The problem of over-focussing of the electron beam by the strong proton-low-$\beta$ quadrupoles is solved using the fact of smallness of the emittance of the TESLA electron beam. For this reason the divergence of the electron beam after conversion will be dominated by the kinematics of the Compton backscattering. In the case of 300 (800) GeV electron beam the maximum value of scattering angle is 4 (1.5) micro-radians. Therefore, the electron beam transverse size will be 100 (37.5) $\mu m$ at the distance of 25 $m$ from conversion region and the focusing quadrupoles for proton beam have negligible influence on the residual electron beam. On the other hand, in the scheme with deflection there is no restriction on $n_e$ from $\Delta Q_p$, therefore, larger $n_e$ and bunch spacing may be preferable. All these topics need a further research.

Concerning the experimental aspects, very forward detector in $\gamma$ beam direction will be very useful for investigation of small $x_g$ region due to registration of charmed and beauty hadrons produced via $\gamma g \rightarrow Q^* Q$ sub-process.

There are a number of papers (see [3] and refs. therein), devoted to physics at $\gamma p$ colliders. Concerning the BSM physics, $\gamma p$ option of TESLA ⊗HERA doesn't promise essential results with possible exclusions of the first family excited quarks (if their masses are less than 1 TeV) and associate production of gaugino and first family squarks (if the sum of their masses are less than 0.5 TeV). The photo-production of $W$ and $Z$ bosons may be also the matter of interest for investigation of the their anomalous couplings. However, $c^* c$ and $b^* b$ pairs will be copiously photo-produced at $x_g$ of order of 10$^{-5}$ and 10$^{-4}$, respectively, and saturation of gluons should manifest itself. Then, there are a number of different photo-production processes (including di-jets etc.) which can be investigated at $\gamma p$ colliders.

### iii) $eA$ option [1,18]

The main limitation for this option comes from fast emittance growth due to intra-beam scattering, which is approximately proportional ($Z^2/A)^2(\gamma_A)^{-3}$. In this case, the use of flat nucleus beams seems to be more advantageous (as in the case of $ep$ option [10]) because of few times increasing of luminosity lifetime. Nevertheless, sufficiently high luminosity can be achieved at least for light nuclei. For example, $L_{eC}$=1.1·10$^{29}$cm$^{-2}$s$^{-1}$ for collisions of 300 GeV energy electrons beam (Table III) and Carbon beam with $n_C$=8·10$^9$ and $\varepsilon_C^N = 1.25\,\pi\cdot mm\cdot mrad$ (rest of parameters as in Table IV). This value corresponds to $L^{int}\cdot A \approx 10\,pb^{-1}$ per working year ($10^7\,s$) needed from the physics point of view [19,20]. Similar to the $ep$ option, the lower limit on $\beta_A^*$, which is given by nucleus bunch length, can be overcome by applying a "dynamic" focusing scheme [9] and an upgrade of the luminosity by a factor 3-4 may be possible.

As mentioned above, the large charge density of nucleus bunch results in strong intra-beam scattering effects and lead to essential reduction of luminosity lifetime ($\approx 1\,h$ for $C$ beam at HERA). There are two possible solutions of this problem for TESLA $\otimes$ HERA. Firstly, one could consider the possibility to re-fill nucleus ring at appropriate rate with necessary modifications of the filling time etc. Alternatively, an effective method of cooling of nucleus beam in main ring should be applied, especially for heavy nuclei. For example, electron cooling of nucleus beams suggested in [21] for $eA$ option of HERA can be used for TESLA $\otimes$ HERA, also.

The basic layout for an $eA$ interaction region can be chosen similar to that for an $ep$ option considered in [7] and [9]. The work on this subject is in progress [18]. A possible layout of a detector for TESLA $\otimes$ HERA based $ep$ collider, which can be used also for $eA$ option, is presented in [13].

The physics search program of the machine is the direct extension of that for $eA$ option of HERA (see Chapter titled "Light and Heavy Nuclei in HERA" in [22]).

## iv)    $\gamma A$ option [1,18]

In my opinion this is a most promising option of TESLA $\otimes$ HERA complex, because it will give unique opportunity to investigate small $x_g$ region in nuclear medium. Indeed, due to the advantage of the real $\gamma$ spectrum heavy quarks will be produced via $\gamma g$ fusion at characteristic

$$x_g \approx \frac{4m^2_{c(b)}}{0.9\times(Z\,/\,A)\times s_{ep}},$$

which is approximately ($2\div3$)$\cdot10^{-5}$ for charmed hadrons. As in the previous option, sufficiently high luminosity can be achieved at least for light nuclei. Then, the scheme with deflection of electron beam after conversion is preferable because it will give opportunity to avoid limitations from $\Delta Q_A$, especially for heavy nuclei. The dependence of luminosity on the distance between conversion region and interaction point for TESLA $\otimes$ HERA based $\gamma C$ collider is similar to that of the $\gamma p$ option [16]: $L_{\gamma C}=2\cdot10^{29}cm^{-2}s^{-1}$ at $z=0$ and $L_{\gamma C}=10^{29}cm^{-2}s^{-1}$ at $z=10\,m$ with 300 GeV energy electron beam. Let me remind you that an upgrade of the luminosity by a factor 3-4 may be possible by applying a "dynamic" focusing scheme. Further increasing of luminosity will require the cooling of nucleus beam in the main ring. Finally, very forward detector in $\gamma$-beam direction will be very useful for investigation of small $x_g$ region due to registration of charmed and beauty hadrons.

Let me finish this section by quoting the paragraph, written taking into mind $eA$ option of the TESLA $\otimes$ HERA complex but more than applicable for $\gamma A$ option, from the recent paper [19]:

"Extension of the $x$-range by two orders of magnitude at TESLA-HERA collider would correspond to an increase of the gluon densities by a factor of 3 for $Q^2=10\,GeV^2$. It will definitely bring quark interactions at this scale into the region where DGLAP will breakdown. For the gluon-induced interactions it would allow the exploration of a non-DGLAP hard dynamics over two orders of magnitude in $x$ in the kinematics where $\alpha_s$ is small while the fluctuations of parton densities are large."

**v) FEL $\gamma A$ option [23,24]**

Colliding of TESLA FEL beam with nucleus bunches from HERA may give a unique possibility to investigate "old" nuclear phenomena in rather unusual conditions. The main idea is very simple [23]: ultra-relativistic ions will see laser photons with energy $\omega_0$ as a beam of photons with energy $2\gamma_A\omega_0$, where $\gamma_A$ is the Lorentz factor of the ion beam. Moreover, since the accelerated nuclei are fully ionized, we will be free from possible background induced by low-shell electrons. For HERA $\gamma_A=(Z/A)\gamma_p=980(Z/A)$, therefore, the region $0.1 \div 10$ MeV, which is matter of interest for nuclear spectroscopy, corresponds to $0.1 \div 10$ keV lasers, which coincide with the energy region of TESLA FEL [4].

The excited nucleus will turn to the ground state at a distance $l=\gamma_A\cdot\tau_A\cdot c$ from the collision point, where $\tau_A$ is the lifetime of the exited state in the nucleus rest frame and $c$ is the speed of light. For example, one has $l=4mm$ for 4438 keV excitation of $^{12}C$. Therefore, the detectors should be placed close to the collision region. The MeV energy photons emitted in the rest frame of the nucleus will be seen in the detector as high-energy photons with energies up to GeV region.

The huge number of expected events (~$10^{10}$ per day for 4438 keV excitation of $^{12}C$) and small energy spread of colliding beams ($\leq 10^{-3}$ for both nucleus and FEL beams) will give opportunity to scan an interesting region with ~1 keV accuracy.

**2.2. LEP $\otimes$ LHC**

The interest in this collider, which was widely discussed [25,26] at earlier stages of LHC proposal, has renewed recently [27]:

"We consider the LHC $e^{\pm}p$ option to be already part of the LHC programme... The availability of $e^{\pm}p$ collisions at an energy roughly four times that provided currently by HERA would allow studies of quark structure down to a size of about $10^{-17}$ cm... The discovery of the quark substructure could explain the Problem of Flavour, or one might discover leptoquarks or squarks as resonances in the direct channel..."

**i) $ep$ option**

The recent set of parameters is given in a report [28] prepared at the request of the CERN Scientific Policy Committee. Parameters of proton beam are presented in corresponding column of Table IV. Parameters of electron beam are as follows: $E_e=67.3$ GeV, $n_e=6.4\cdot10^{10}$, $\varepsilon_x/\varepsilon_y=9.5/2.9$ nm and $\beta_x^*/\beta_y^*=85/26cm$. With these parameters the estimated luminosity is $L_{ep}=1.2\cdot10^{32}$ cm$^{-2}$s$^{-1}$ and exceeds that of the HERA $\otimes$ TESLA based $ep$ collider. However, the latter has the advantage in kinematics because of comparable values of energies of colliding particles. Moreover, $\gamma p$ collider cannot be constructed on the base of LEP $\otimes$ LHC (for reasons see [16]).

**ii) $eA$ option**

An estimation of the luminosity for $e^{\pm}Pb$ collisions given in [28] seem to be over-optimistic because of unacceptable value of $\Delta Q_{Pb}$, which is ~$70 \cdot 10^{-3}$ for proposed set of parameters. The one order lower value $L_{e-Pb}=10^{28}$cm$^{-2}$s$^{-1}$ is more realistic. However, situation may be different for light nuclei. Again the main advantage of the TESLA⊗HERA complex in comparison with LEP ⊗LHC is $\gamma A$ option.

### 2.3. $\mu-ring$⊗TEVATRON [29]

If the main problems (high $\mu$–production rate, fast cooling of $\mu$–beam etc.) facing $\mu^+\mu^-$ proposals are successfully solved, it will also give an opportunity to construct $\mu p$ colliders. Today only very rough estimations of the parameters of these machines can be made. Two sets of parameters for the collider with two rings, TEVATRON and 200 GeV muon ring, are considered in [29]. Possible parameters of muon beam are presented in Table V and upgraded parameters of TEVATRON proton beam are given in the Table IV (numbers in brackets correspond to option with low $\mu$–production rate). Let us mention the large values of proton beam tune in the first option and proton bunch population in the second option. In my opinion, luminosity values presented in [29] are over-optimistic. For comparison, recent set of 200 GeV muon beam parameters [30], given in the Table V (symbol $\rightarrow$ means transition from 1000 $m$ circumference to TEVATRON), and corresponding parameters from Table IV (proton beam emittance is estimated from $\Delta Q_p$=0.003) lead to estimation $L_{\mu p}$=1.7·10$^{31}$cm$^{-2}$s$^{-1}$. Nevertheless, with using a larger number of bunches and applying "dynamic" focusing scheme [9], the value of $L_{\mu p}$ order of 10$^{32}$cm$^{-2}$s$^{-1}$ seem to be achievable.
Physics search program of this machine is similar to that of the $ep$ option of the TESLA⊗HERA complex.

## 3. Second stage: Linac ⊗LHC and $\sqrt{s}$=3TeV $\mu p$

### 3.1. Linac ⊗LHC

The center-of-mass energies which will be achieved at different options of this machine are an order larger than those at HERA are and ~3 times larger than the energy region of TESLA ⊗HERA, LEP ⊗LHC and $\mu-ring$⊗TEVATRON. In principle, luminosity values are ~7 times higher than those of corresponding options of TESLA⊗HERA complex due to higher energy of protons. Following [7,8] below we consider electron linac with P$_e$=60 MW and upgraded proton beam parameters given in the last column of the Table IV.

### i) $ep$ option [7,8,15,31]

According [7, 8] center-of-mass energy and luminosity for this option are (with obvious modification due to re-scale of $E_p$ from 8 TeV to 7 TeV): $\sqrt{s}$ =5.29 TeV and $L_{ep}$=8·10$^{31}$cm$^{-2}$s$^{-1}$, respectively. Let me remind you that an additional factor 3-4 can be provided by the "dynamic" focusing scheme. Further increasing will require cooling at injector stages.

 This machine, which will extend both the $Q^2$-range and $x$-range by more than two order of magnitude comparing to those explored by HERA, has a strong potential for both SM and BSM research.

**ii)** $\gamma p$ **option** [15,16]

The advantage in spectrum of back-scattered photons (for details see ref. [16]) and sufficiently high luminosity ($L_{\gamma p} > 10^{32}$cm$^{-2}$s$^{-1}$) will clearly manifest itself in a search for different phenomena. For example, thousands di-jets with $p_t > 500$ GeV and hundreds thousands single $W$ bosons will be produced, hundred millions of $b^*b$- and $c^*c$-pairs will give opportunity to explore the region of extremely small $x_g$ etc.

**iii)** $eA$ **option** [1,18,32]

In the case of LHC nucleus beam IBS effects in main ring are not crucial because of larger value of $\gamma_A$. The main principal limitation for heavy nuclei coming from beam-beam tune shift may be weakened using flat beams at collision point. Rough estimations show that $L_{eA} A > 10^{31}$cm$^{-2}$s$^{-1}$ can be achieved at least for light and medium nuclei.

**iv)** $\gamma A$ **option** [1,18,32]

Limitation on luminosity due to beam-beam tune shift is removed in the scheme with deflection of electron beam after conversion.

The physics search potential of this option, as well as that of previous three options, needs more investigations from both particle and nuclear physics viewpoints.

**v) FEL** $\gamma A$ **option** [23,24]

Due to a larger $\gamma_A$, the requirement on wavelength of the FEL photons is weaker than in the case of TESLA $\otimes$ HERA based FEL $\gamma A$ collider. Therefore, the possibility of constructing a special FEL for this option may be m atter of interest. In any case the realization of FEL $\gamma A$ colliders depends on the position of "traditional" nuclear physics community.

**3.2.** $\sqrt{s} = 4(\rightarrow 3)$ **TeV** $\mu p$ [33]

The possible $\mu p$ collider with $\sqrt{s} = 4$ TeV in the framework of $\mu^+\mu^-$ project was discussed in [33] and again over-estimated value of luminosity, namely, L$_{\mu p} = 3\cdot10^{35}$cm$^{-2}$s$^{-1}$, was considered. Using recent set of parameters [30] for high-energy muon collider with $\sqrt{s} = 3$ TeV, one can easily estimate possible parameters of $\mu p$ collisions from:

$$L_{\mu p} = \frac{n_p}{n_\mu} \cdot \frac{\beta_\mu^*}{\beta_p^*} \cdot \frac{m_\mu}{m_p} \cdot \frac{\varepsilon_\mu^N}{\varepsilon_p^N} \cdot L_{\mu^+\mu^-} .$$

With $n_p = n_\mu = 2\cdot10^{12}$ we obtain (normalized emittance of the proton beam is estimated from $\Delta Q_p = 0.003$)

$$L_{\mu p} = \frac{0.3cm}{15cm} \cdot \frac{105MeV}{940MeV} \cdot \frac{50}{80} \cdot 7\cdot10^{34}\ cm^{-2}s^{-1} = 10^{32}\ cm^{-2}s^{-1} .$$

In principle, an upgrade of the luminosity by a fac tor 3-4 may be possible by applying a "dynamic" focusing scheme.

This machine is comparable with the $ep$ option of the Linac $\otimes$ LHC.

# 4. Third stage: $e-ring \otimes VLHC$, LSC $\otimes ELOISATRON$ and 100 TeV $\mu p$

## 4.1. $e-ring \otimes VLHC$

Over the last decade two projects, ELOISATRON [34] and VLHC [35] have been intensively discussed for proton-proton collisions at 100 TeV energy range. The possible installation of a large electron/positron ring (top-factory) in VLHC tunnel will give an opportunity to get $ep$ collisions with $\sqrt{s} = 7$ TeV [36]. Recently, an $ep$ collider with $\sqrt{s} = 1$ TeV ($E_e = 80$ GeV, $E_p = 1$ TeV) and $L_{ep} = 3.2 \cdot 10^{32}$ cm$^{-2}$s$^{-1}$ in a VLHC Booster Tunnel with ~34 km circumference has been proposed [36, 37]. Le t me cite the Conclusions from [37]:

"We have done a preliminary study of an $ep$ collider that could be installed in the low field booster of the VLHC. This machine could be op erational before the LEP/LHC and would have a higher luminosity than HERA/TESLA."

## 4.2. LSC $\otimes$ ELOISATRON

For obvious reasons I prefer to discuss a lin ac, e.g. Linear Super Collider [38], as a source of high-energy electron beam. Combination of LSC with ELOISATRON (see Table VI) will give an opportunity to achieve $L_{ep} = 3 \cdot 10^{32}$ cm$^{-2}$s$^{-1}$ at $\sqrt{s} = 63.2$ TeV. Further increasing of luminosity will require application o f "dynamic" focusing scheme and/or cooling of proton beam.

As in the case of TESLA $\otimes$ HERA complex, $\gamma p$, $eA$, $\gamma A$ and FEL $\gamma A$ options essentially extend the capacity of LSC $\otimes$ ELOISATRON complex.

## 4.3. 100 TeV $\mu p$

This is a most speculative (however, very att ractive) one among the lepton-hadron collider options, which can be foreseen today. Usin g the luminosity estimation $L_{\mu\mu} = 10^{36}$ cm$^{-2}$s$^{-1}$ for $\sqrt{s} = 100$ TeV $\mu^+\mu^-$ collider [39] one can expect at $n_p = n_\mu$

$$L_{\mu p} \approx \frac{0.25 cm}{10 cm} \cdot \frac{105 MeV}{940 MeV} \cdot \frac{8.7}{30} L_{\mu\mu} \approx 10^{33} cm^{-2} s^{-1}.$$

# 5. Conclusion

It seems that neither HERA nor LHC $\otimes$ LEP will be the end points for lepton-hadron colliders. Linac-ring type $ep$ machines and possibly $\mu p$ colliders will give an opportunity to go far in this direction (see Table VII). Howeve r, more activity is needed both in accelerator (further exploration of "dynamic" focus ing scheme, a search for effective cooling methods etc.) and physics search program as pects.

# Acknowledgements

I would like to express my gratitude to DESY Directorate for invitation and hospitality. I am grateful to D. Barber, V. Borodul in, R. Brinkmann, O. Cakir, A. Celikel, A. K. Ciftci, L. Frankfurt, P. Handel, M. Klein, M. Leenen, C. Niebuhr, M. Strikman, V. Telnov, D. Trines, G. A. Voss, A. Wagner, F. Willek e and O. Yavas for useful and stimulating discussions. Special thanks are to N. W alker for careful reading of the manuscript and valuable comments. My work on the su bject was strongly encouraged by the support of Professor B.Wiik, who personally visited Ankara in 1996 and signed the Collaboration Agreement between DESY and Ankara University. This work is supported in part by Turkish State Planning Organization under the grant number 97K-120-420.

Table I. Parameters of electron beams for $\sqrt{s}$ =500 GeV linear colliders

| | TESLA | JLC(C) | JLC/NLC | CLIC |
|---|---|---|---|---|
| Linac repetition rate (Hz) | 5 | 100 | 120 | 200 |
| No. of particles/bunch (10$^{10}$) | 2 | 1.11 | 0.95 | 0.4 |
| No. of bunches/pulse | 2820 | 72 | 95 | 150 |
| Bunch separation (ns) | 337 | 2.8 | 2.8 | 0.67 |
| Beam power (MW) | 11.3 | 3.07 | 4.5 | 4.81 |
| $\gamma\varepsilon_x/\gamma\varepsilon_y$ (mm·mrad) | 10/0.03 | 3.3/0.05 | 4.5/0.1 | 1.88/0.1 |
| $\beta_x^*/\beta_y^*$ (mm) | 15/0.4 | 15/0.2 | 12/0.12 | 10/0.1 |
| $\sigma_x/\sigma_y$ at IP (nm) | 535/5 | 318/4.3 | 330/4.9 | 196/4.5 |
| $\sigma_z$ (µm) | 400 | 200 | 120 | 50 |

Table II. Parameters of proton beams

| | HERA | TEVATRON | LHC |
|---|---|---|---|
| Beam energy (TeV) | 0.92 | 1 | 7 |
| No. of particles/bunch (10$^{10}$) | 7 | 27 | 10.5 |
| No. of bunches/ring | 180 | 36 | 2835 |
| Bunch separation (ns) | 96 | 396 | 25 |
| $\varepsilon$ (10$^{-9}$ $\pi$rad·m) | 5 | 3.5 | 0.5 |
| $\beta_x^*/\beta_y^*$ (m) | 7/0.5 | 0.35/0.35 | 0.5/0.5 |
| $\sigma_x/\sigma_y$ at IP (µm) | 265/50 | 34/34 | 16/16 |
| $\sigma_z$ (cm) | 8.5 | 38 | 7.5 |

Table III. Upgraded parameters of the TESLA electron beams

| Electron energy, GeV | 300 | 500 | 800 |
|---|---|---|---|
| Number of electrons per bunch, 10$^{10}$ | 1.8 | 2→1.5 | 1.41 |
| Number of bunches per pulse | 4640 | 2820→4950 | 4500→4430 |
| Beam power, MW | 40 | 22.6→30 | 24.4→40 |
| Bunch length, mm | 1 | 0.4 | 0.3 |
| Bunch spacing, ns | 192 | 337→192 | 189→192 |
| Repetition rate, Hz | 10 | 5 | 3→5 |

TableIV.Upgradedparametersofprotonbeams

| | TESLA⊗HERA | LEP⊗LHC | $\mu p$ TEVATRON | Linac ⊗LHC |
|---|---|---|---|---|
| No of protons per bunch,$10^{10}$ | 30 | 10 | 125(500) →200 | 30 |
| No of bunches per ring | 90 | 508 | 2→4 | ~800 |
| Bunch spacing,ns | 192 | 100 | 7128 | 100 |
| $\varepsilon_p^N$,$10^{-6}$ m | 0.8 | 3.75 | 12.5(50) →80 | 0.8 |
| $\sigma_z$,cm | 15 | 7.5 | 15 | 7.5 |
| $\beta_p^*$,cm | 20 | 16/1.5m | 15 | 10 |
| $\sigma_{x,y}$ at IP,  μm | 12.5 | 90/28 | 40(90) →106 | 3.3 |
| $\Delta Q_p$,$10^{-4}$ | 27 | 32/10 | 200(5) →30 | 20 |

TableV.Parametersofmuonbeams

| | $\mu p$ high(low) | Recentset | Resentset |
|---|---|---|---|
| Energy,GeV | 200 | 200 | 1500 |
| Number of muons per bunch,$10^{10}$ | 200(50) | 200 | 200 |
| Number of bunches per ring | 2 | 4 | 4 |
| Number of turns per pulse | 2000 | 700→100 | 785 |
| Pulse rate,Hz | 30(10) | 15 | 15 |
| $\varepsilon_\mu^N$,$10^{-6}$ m | 50(200) | 50 | 50 |
| $\beta_\mu^*$,cm | 7.5 | 2.6 | 0.3 |

Table VI. Parameters of the future multi-TeV beams

| Particles | p(ELOISATRON) | e$^{\pm}$(LSC) | μ$^{\pm}$( $\sqrt{s}$ =100TeV  μ$^+$μ$^-$) |
|---|---|---|---|
| Energy,TeV | 100 | 10 | 50 |
| Number of particles per bunch,10$^{10}$ | 0.51→30 | 0.041 | 80 |
| Number of bunches per pulse | - | 1 | 1 |
| Number of bunches per ring | 43960→22 | - | 1 |
| Bunch separation,m | 5→10000 | 10000 | 100000 |
| Repetition rate,Hz | - | 30000 | 7.9 |
| Number of turns | - | - | 1350 |
| $\varepsilon^N$,10$^{-6}$m | 0.9 | 0.04 | 8.7 |
| σ$_{x,y}$,10$^{-6}$m | 0.67 →0.3 | 0.002 | 0.21 |
| $\beta^*$,cm | 50 →10 | 0.1 | 0.25 |

Table VII. Future lepton-hadron colliders:

a) First stage (2010-2015)

| | TESLA⊗HERA | LEP ⊗LHC | μ-ring⊗Tevatron |
|---|---|---|---|
| $\sqrt{s}$ ,TeV | 1.05(1.35)  →1.79 | 1.37 | 0.89 |
| $E_l$,TeV | 0.3(0.5) →0.8 | 0.0673 | 0.2 |
| $E_p$,TeV | 0.92 →1 | 7 | 1 |
| $L$,10$^{31}$cm$^{-2}$s$^{-1}$ | 1÷10 | 12 | 1÷10 |
| Main limitations | $P_e$, $\varepsilon_p$, $\beta_p^*$ | $\Delta Q_e$, $\Delta Q_p$ | $n_\mu$, $\varepsilon_\mu$, $\Delta Q_p$, $\varepsilon_p$, $\beta_p^*$ |
| Additional options | $eA$, $\gamma p$, $\gamma A$,FEL  $\gamma A$ | $eA$ | $\mu A$(?) |

b) Second (2015-2020) and third (>2020) stages

| | Linac⊗LHC | $\mu p$ | LSC ⊗ELOISATRON | $\mu p$ |
|---|---|---|---|---|
| $\sqrt{s}$ ,TeV | 5.29 | 3 | 63.2 | 100 |
| $E_l$,TeV | 1 | 1.5 | 10 | 50 |
| $E_p$,TeV | 7 | 1.5 | 100 | 50 |
| $L$,10$^{31}$cm$^{-2}$s$^{-1}$ | 10÷100 | 10 ÷100 | 100 ÷1000 | 100 ÷1000 |
| Options | $eA$, $\gamma p$, $\gamma A$,FEL  $\gamma A$ | $\mu A$(?) | $eA$, $\gamma p$, $\gamma A$,FEL  $\gamma A$ | $\mu A$(?) |